\documentclass[a4paper,twocolumn,english,prl,superscriptaddress,showpacs,tight]{revtex4}
\usepackage{ae,aecompl}
\usepackage[T1]{fontenc}
\usepackage[latin1]{inputenc}
\usepackage{float}
\usepackage{amsmath}
\usepackage{graphicx}
\usepackage{amssymb}

\makeatletter

\providecommand{\tabularnewline}{\\}

\@ifundefined{textcolor}{}
{%
 \definecolor{BLACK}{gray}{0}
 \definecolor{WHITE}{gray}{1}
 \definecolor{RED}{rgb}{1,0,0}
 \definecolor{GREEN}{rgb}{0,1,0}
 \definecolor{BLUE}{rgb}{0,0,1}
 \definecolor{CYAN}{cmyk}{1,0,0,0}
 \definecolor{MAGENTA}{cmyk}{0,1,0,0}
 \definecolor{YELLOW}{cmyk}{0,0,1,0}
 }

\usepackage{epsfig}

\usepackage{bm}

\makeatother

\usepackage{babel}

\makeatother

\usepackage{babel}

\makeatother

\usepackage{babel}

\begin{document}

\title{Emergence of robust gaps in 2D antiferromagnets via additional spin-1/2
probes}

\author{Aires Ferreira, J. Viana Lopes, and J. M. B. Lopes dos Santos}

\affiliation{CFP and Departamento de Física, Faculdade Ciências Universidade do
Porto, 687 4169-007, Porto, Portugal}
\begin{abstract}
We study the capacity of antiferromagnetic lattices of varying geometries
to entangle two additional spin-$1/2$ probes. Analytical modeling
of the Quantum Monte Carlo data shows the appearance of a robust gap,
allowing a description of entanglement in terms of probe-only states,
even in cases where the coupling to the probes is larger than the
gap of the spin lattice and cannot be treated perturbatively. We find
a considerable enhancement of the temperature at which probe entanglement
disappears as we vary the geometry of the bus and the coupling to
the probes. In particular, the square Heisenberg antiferromagnet exhibits
the best thermal robustness of all systems, whereas the three-leg
ladder chain shows the best performance in the natural quantum ground
state. 
\end{abstract}
\maketitle

\section*{I. INTRODUCTION}

Solid-state systems have been exploited to integrate quantum information
(QI) tasks and accomplish quantum computation in a single processing
core, but many questions regarding their robustness against temperature
and decoherence remain open. Among the requirements to achieve quantum
computation, the ability to generate rapid elementary gates between
well-characterized qubits is central \cite{DiVicenzo}. In view of
the technical difficulties of switching on direct interactions between
qubits, various proposals have been put forward to use a quantum sub-system,
usually denominated as \emph{bus,} to mediate the fundamental universal
gates \cite{Cirac,DiVincenzo2}. A considerable body of work has been
devoted to chains of spins experiencing nearest-neighbor interactions,
since they can be used as models for universal quantum computation,
meeting the aforementioned requirements \cite{SolidStateQC_withBus}.

Spin chains are indeed a versatile laboratory for QI science for they
naturally embody the SU$(2)$ algebra of a qubit, allowing quantum
information processing and manipulation along the traditional lines
of quantum computing, that is, via the establishment of quantum gates
\cite{DiVicenzo}. In particular, numerical simulations \cite{Venuti}
showed that spin systems can mediate entanglement between two spin
probes separated by large distances, the so-called long-distance entanglement
(LDE)---a quite remarkable phenomenon since, away from critical points,
bulk correlations are completely classical at distances larger than
a few sites \cite{Osborne}; gapless bosonic systems do not display
this long distance entanglement capability \cite{Zell09}.

The possibility of entangling spin systems via interaction with a
larger system of spins was first pointed out by Chiara et al. in the
proposal for entanglement extraction from solids \cite{DeChiara}.
Entanglement extraction from a large system might seem nonintuitive
as generally the coupling to a system with many degrees of freedom
leads unavoidably to decoherence \cite{Review-Zureck-2003}. However,
a few notable exceptions are known: If two qubits, not interacting
directly, are coupled in a symmetric way to a bath of harmonic oscillators,
their entanglement will partially survive during their evolution when
these qubits have degenerate energy eigenstates \cite{2002Braun,2003Benatti,2008Paz}
or when the bath has a gap in its spectrum \cite{2006Oh}. This is
reminiscent of quantum computing using the so-called decoherence-free
subspaces \cite{1998Lidar,2000Beige}.

\begin{figure}
\noindent \begin{centering}
\includegraphics[width=0.7\columnwidth]{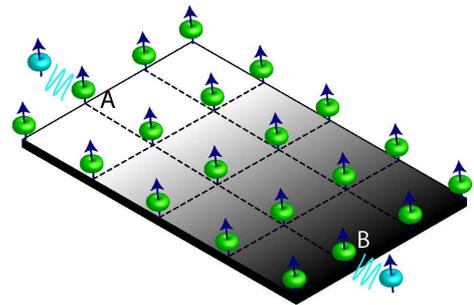} 
\par\end{centering}

\caption{\label{fig:Chap5-Lattice}Schematic of the total system: The probes
(blue) interact locally with a 2D lattice (the bus) through sites
$A$ and $B$. It is well known that spin chains can mediate entanglement
between additional spin-$1/2$ particles, even at large distances
\cite{Venuti,Venuti-b,AFF_LDS} (the so-called LDE), but the question
of how increasing dimensionality changes the 1D LDE picture has not
been addressed until now.}

\end{figure}

Promising advances in the engineering of atomic structures and optical
lattices, where finite spin systems are effectively realized in the
laboratory, encourage the consideration of more general possibilities.
In this article, we study the effect of adding two spin-$1/2$ probes
to various antiferromagnetic spin-$1/2$ lattices, ranging from a
one-dimensional chain to a two-dimensional (2D) antiferromagnetic
(AF) lattice (the bus in QI terminology), on the probes LDE and spectrum.
We confirm previous numerical results for the one-dimensional scenario
\cite{Venuti-b} and extend this analysis to two dimensions via the
first \emph{direct} measure of probe correlation, with quantum Monte
Carlo (QMC) methods. Our main findings are (1) an enhancement of LDE
robustness regarding temperature in 2D AF Heisenberg systems, relative
to the 1D chain; (2) the observation of a significant entanglement
of the distant probes, even when the coupling to the bus is too strong
to be described as a weak perturbation of the bus; and (3) the possibility
of a complete description of the temperature dependence of the entanglement
of the probes in terms of a low energy manifold of singlet and triplet
levels, with a large singlet-triplet gap, protecting probe entanglement
from the undesirable effects of temperature---an unexpected phenomenon,
especially for the square AF spin system, which \emph{per se} has
a vanishing gap and a broken symmetry phase in the thermodynamic limit
\cite{Review-Manousakis}. This is a situation favorable to QI capabilities,
particularly in the case of a 2D spin bus, where a relatively large
singlet-triplet emerges with increasing probe-bus coupling, the singlet
retaining significant probe-probe entanglement despite the stronger
probe-bus coupling; we refer to this situation as presenting a {}``robust
gap.''

The article is organized as follows. We start by making precise the
conditions for a \emph{robust gap} and set up the analytical theory
which will allow us to conclude about the nature of the spectrum from
the numerical simulations. The basic notions on LDE will also be reviewed
(Sec. II). In Sec. III, the numerical results will be presented; in
particular, (1) the singlet-triplet gap and LDE for all measured systems,
(2) the robustness of thermal LDE in 2D, and (3) a surprisingly efficient
entangler bus at $T=0$, the three-leg ladder. We finish with conclusions
and a discussion of consequences of the present work for LDE (Sec.
IV). Details of derivations are presented in the appendices.

\section*{II. Analytical model}

In this section, we introduce the analytical tools with which we modeled
the Monte Carlo data. A few analytical results will be derived under
the assumption of emergence of \emph{a robust gap.} By this we mean
the possibility of completely describing\emph{ }the probe's entanglement,
and its temperature variation, in terms of a low-lying singlet-triplet
manifold which is isomorphic to the probe's space of states. An analytical
function, containing three parameters which are, in principle, calculable
perturbatively, was found to fit the Monte Carlo data for probe entanglement
almost perfectly.

\subsection*{A. Adiabatic continuity\emph{ }}

We admit global SU$(2)$ symmetry of the couplings; more general types
of interactions could be easily considered within this framework,
but we focus on Heisenberg interactions, which not only allow universal
quantum computation \cite{DiVincenzo2} but are also commonly realized
in nature (e.g., in the parent compounds of copper-oxide high-temperature
superconductors such as the undoped insulator La$_{2}$CuO$_{4}$
\cite{Review-Manousakis}; in electronically coupled quasi-1D chains
such as the CuGeO$\textrm{\ensuremath{_{3}}}$ \cite{CuGeO3}; in
the Mott insulating one-dimensional perovskite KCuF$_{3}$ \cite{1D-KCuF3};
and in linear chains of $\sim10$ manganese atoms in engineered atomic
structures \cite{AtomicChains}). We start by considering many-body
systems of spins (which we will designate by bus) with rotational
invariant Hamiltonian, $H_{0}$, and a singlet, nondegenerate, ground
state. Two probes, $\boldsymbol{\tau}_{a}$ and $\boldsymbol{\tau}_{b}$
($\boldsymbol{\tau}$ being the Pauli matrices), are coupled to the
bus by Heisenberg exchange interaction with strength $J_{p}:=J\alpha$,
through sites $A$ and $B$, respectively, \begin{equation}
V=\alpha J\left(\vec{\boldsymbol{S}}_{A}\cdot\vec{\boldsymbol{\tau}}_{a}+\vec{\boldsymbol{S}}_{B}\cdot\vec{\boldsymbol{\tau}}_{b}\right),\label{eq:Chap5-Eq1}\end{equation}
where $J$ denotes an energy scale of the bus (typically its exchange
interaction), $\boldsymbol{S}_{A(B)}=(1/2)\boldsymbol{\tau}_{A(B)}$,
and $\alpha$ is a dimensionless parameter. We make the simple but
crucial assumption that there is a one-to-one map of eigenstates of
the uncoupled system ($\alpha=0$) to the eigenstates of the full
Hamiltonian; that is, we invoke adiabatic continuity \cite{Anderson}.
Hence we define a canonical transformation between the two basis:\begin{equation}
|\psi_{m}\rangle\otimes|\chi_{\sigma}^{ab}\rangle=e^{-i\hat{S}}|\Psi_{m,\sigma}\rangle,\label{eq:Chap5-Eq2}\end{equation}
 where $|\psi_{m}\rangle$ is a bus-only eigenstate, $|\chi_{\sigma}^{ab}\rangle$
a probe state, and $|\Psi_{m,\sigma}\rangle$ an eigenstate for finite
$\alpha$. Note that the generator $\hat{S}$ is an operator acting
on both probe and bus space. This map has important consequences:
The transformed Hamiltonian must have the form of a sum of a probe-only
term ($H_{p}$) with a bus-only term ($H_{b}^{'}$), that is, $H\rightarrow H_{p}+H_{b}^{\prime}$,
\begin{figure}
\noindent \centering{}\includegraphics[width=1\columnwidth]{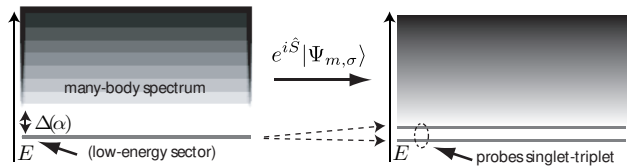}\caption[The canonical transformed spectrum under the robust gap assumption]{\label{fig:Chap5-CanonicalTransf}Schematic of the canonical transformed
many-body spectrum (at right) under the assumptions of adiabatic continuity
and a finite gap $\Delta(\alpha)$. If one succeeds in finding the
matrix elements of $\hat{S}$, the low-energy physics of our problem
will be described by an effective Hamiltonian containing just the
probe's canonical singlet and triplet states. The transformation $\hat{S}$
renormalizes both operators and states.}

\end{figure}
since the corresponding eigenstates are product states. We now add
the assumption that the lowest lying states which map to a probe singlet
and probe triplet, \begin{eqnarray}
|\Psi_{0,s}\rangle & = & e^{i\hat{S}}|\psi_{0}\rangle\otimes|\chi^{s}\rangle,\label{eq:Chap5-Eq4}\\
|\Psi_{0,m}\rangle & = & e^{i\hat{S}}|\psi_{0}\rangle\otimes|\chi_{m}^{t}\rangle\qquad m=0,\pm1,\label{eq:Chap5-Eq5}\end{eqnarray}
respectively, are well-separated from states which map to excited
states of the bus by a finite gap, $\Delta(\alpha)$. In the scenario
we have in mind---and which is confirmed by the simulations---entanglement
between the probes disappears well before significant thermal occupation
of states above $\Delta(\alpha)$ occurs. But an important remark
is in order: While the probes are obviously entangled in the state
$|\chi^{s}\rangle$, it is the entanglement in the state $|\Psi_{0,s}\rangle$
which matters. In other words, the canonical transformation changes
operators as well as states; if we are to use the transformed uncoupled
basis in the calculation, the probe entanglement \emph{is }not\emph{
}the entanglement of $\boldsymbol{\tau}_{a}$ and $\boldsymbol{\tau}_{b}$,
but rather the entanglement of the renormalized spins\emph{,} \begin{equation}
\boldsymbol{\tau}_{a(b)}^{R}:=e^{-i\hat{S}}\boldsymbol{\tau}_{a(b)}e^{i\hat{S}}.\label{eq:Chap5-RenormalizedSpins}\end{equation}
We now proceed to use this concept of adiabatic continuity to develop
a description of the probe-reduced density matrix.

\subsection*{B. Robust gap and canonical corrections.}

The exact density matrix of the qubits (the partial state of the probes,
$\rho_{ab}$) encodes the full capabilities of a generic lattice as
a quantum bus (in particular, the possibility of LDE) and is given
by the Gibbs canonical state\begin{equation}
\rho_{ab}=\mathcal{Z}_{ab}^{-1}\textrm{Tr}_{i\in\mathcal{L}}\left[e^{-\beta\left(H_{0}+V\right)}\right].\label{eq:PartialState - Def}\end{equation}
 The trace is made with respect to the degrees of freedom of the bus,
$\mathcal{L}$, and $\mathcal{Z}_{ab}=\textrm{Tr}[\exp(-\beta H_{0}-\beta V)]$
is the system's partition function. In our case, global $SU(2)$ symmetry
implies a very simple form for $\rho_{ab}$,\begin{equation}
\rho_{ab}\propto\exp\left[-\beta J_{ab}(\beta)\boldsymbol{\tau}_{a}\cdot\boldsymbol{\tau}_{b}\right],\label{eq:rho_ab}\end{equation}
 where $J_{ab}(\beta)$ is the actual effective coupling of the probes
and depends upon the temperature, as a consequence of tracing out
the degrees of freedom of $\mathcal{L}$. If this function is known,
bipartite entanglement can be computed using the negativity \cite{negativity},
the concurrence \cite{concurrence}, or any other entanglement monotone.
For systems with full rotational symmetry entanglement, bipartite
entanglement {[}$E(\rho_{ab})${]} is a function of the probe correlation,
\begin{equation}
\langle\boldsymbol{\tau}_{a}\cdot\boldsymbol{\tau}_{b}\rangle=\textrm{Tr}\left[\rho_{ab}\boldsymbol{\tau}_{a}\cdot\boldsymbol{\tau}_{b}\right]=\frac{e^{-4\beta J_{ab}(\beta)}-1}{e^{-4\beta J_{ab}(\beta)}+1/3}.\label{eq:tau_atau_b}\end{equation}
 The concurrence $E_{C}$ is of particular simplicity for rotational
$2\otimes2$ systems,\begin{equation}
E_{C}(\rho_{ab})=\max\left[0,\left|\frac{\langle\boldsymbol{\tau}_{a}\cdot\boldsymbol{\tau}_{b}\rangle}{3}\right|-\frac{\langle\boldsymbol{\tau}_{a}\cdot\boldsymbol{\tau}_{b}\rangle+3}{6}\right]\in[0,1].\label{eq:Chap5-Concurrence}\end{equation}
 The probes will be entangled whenever $E_{C}(\rho_{ab})>0$, which
happens for strong AF correlations, more precisely, when $\langle\boldsymbol{\tau}_{a}\cdot\boldsymbol{\tau}_{b}\rangle<-1$.
One straightforwardly derives the critical temperature, $T_{c}\equiv1/\beta_{c}$,
above which entanglement vanishes for the state (\ref{eq:rho_ab}).
It reads, \begin{equation}
\beta_{c}J_{ab}(\beta_{c})\simeq0.27.\label{eq:critical temperature}\end{equation}
 One speaks about long-distance entanglement whenever $E(\rho_{ab})>0$
for probes distances of the order of the system size. In the remainder
of this section, we present the analytical form of $J_{ab}(\beta)$,
under the assumption of a robust gap.

In \cite{AFF_LDS}, the authors derived the effective probe interaction
$H_{\textrm{eff}}$ for probes weakly coupled to a large system with
a gap to the first excited state, $\Delta_{\textrm{0}}\equiv\Delta(\alpha=0)>0$.
For $SU(2)$ symmetric couplings, it reads as follows,\begin{equation}
H_{\textrm{eff}}=\alpha^{2}J^{2}\tilde{\chi}_{r}(0)\boldsymbol{\tau}_{a}\cdot\boldsymbol{\tau}_{b},\label{eq:effective_hamiltonian}\end{equation}
where $\tilde{\chi}_{r}(0)$ is the Fourier transform at zero frequency
of the adiabatic spin susceptibility defined as $\chi_{r}(t)=-i\langle[S_{0}^{z}(t),S_{r}^{z}]\rangle\theta(t)$,
with $\theta(t)$ being the Heaviside function and $r$ the distance
between spins $A$ and $B$ in the lattice. The effective Hamiltonian
(\ref{eq:effective_hamiltonian}) holds strictly when, \begin{equation}
J_{\textrm{eff}}:=\alpha^{2}J^{2}\tilde{\chi}_{r}(0)\ll\Delta_{0},\label{eq:J_can_definition_and_condition}\end{equation}
and correctly describes the low-lying spectrum in that limit. However,
as remarked previously, one should be careful in using this Hamiltonian
to calculate probe correlations \cite{AFF_LDS}. Consider, for instance,
the zero-temperature limit. Assuming a system with AF correlations
{[}$\tilde{\chi}_{r}(0)>0${]}, according to Eq.~(\ref{eq:effective_hamiltonian}),
we have, \begin{equation}
\langle\boldsymbol{\tau}_{a}\cdot\boldsymbol{\tau}_{b}\rangle=\frac{1}{\mathcal{Z}_{ab}}\textrm{Tr}\left[\boldsymbol{\tau}_{a}\cdot\boldsymbol{\tau}_{b}e^{-\beta\alpha^{2}J^{2}\tilde{\chi}_{r}(0)\boldsymbol{\tau}_{a}\cdot\boldsymbol{\tau}_{b}}\right]\underset{\beta\rightarrow\infty}{\rightarrow}-3,\label{eq:Chap5-Approximation}\end{equation}
in which case the probe partial state is a perfect singlet with zero
linear entropy: $S_{L}(\rho_{ab}):=1-\textrm{Tr}[\rho_{ab}^{2}]=0$.
This result does not depend on the accuracy of the perturbative estimate
of $J_{\textrm{eff}}$, for we would obtain the same value provided
that $J_{\textrm{eff}}>0$ is temperature independent. On the other
hand, the probes are also correlated (albeit weakly) with the bus,
since the effective probe coupling is a result of perturbative admixture
of virtual excited bus states in the ground-state manifold; the probe
entanglement, even at zero temperature, is not complete.

This issue is clarified by computing the connected correlation between
the probes under the adiabatic assumption and comparing it with (\ref{eq:Chap5-Approximation})
{[}obtained from the approximation, $J_{ab}(\beta)\approx J_{\textrm{eff}}${]}.
This is achieved by recasting the correlation {[}Eq.~(\ref{eq:tau_atau_b}){]}
into a form involving the nonperturbative effective Hamiltonian $H_{S}=H_{b}^{\prime}+H_{p}$.
To this end, we make use of the canonical transformation {[}Eq.~(\ref{eq:Chap5-Eq2}){]}
to get \begin{equation}
\langle\boldsymbol{\tau}_{a}\cdot\boldsymbol{\tau}_{b}\rangle=\frac{\textrm{Tr}\left[e^{-\beta H_{S}}\left(e^{-i\hat{S}}\boldsymbol{\tau}_{a}\cdot\boldsymbol{\tau}_{b}e^{i\hat{S}}\right)\right]}{\textrm{Tr}\left[e^{-\beta H_{S}}\right]}.\label{eq:Chap5-Eq6}\end{equation}
 From the adiabatic continuity assumption {[}Eqs.~(\ref{eq:Chap5-Eq4})-(\ref{eq:Chap5-Eq5}){]},
we get\begin{eqnarray}
\langle\boldsymbol{\tau}_{a}\cdot\boldsymbol{\tau}_{b}\rangle & = & \frac{\textrm{Tr}_{p}\left[e^{-\beta H_{p}}\left(\textrm{Tr}_{b}e^{-\beta H_{b}}e^{-i\hat{S}}\boldsymbol{\tau}_{a}\cdot\boldsymbol{\tau}_{b}e^{i\hat{S}}\right)\right]}{\textrm{Tr}_{p}\left[e^{-\beta H_{p}}\right]\textrm{Tr}_{b}\left[e^{-\beta H_{b}}\right]}\label{eq:6b}\\
 & = & \frac{\textrm{Tr}_{p}\left[e^{-\beta H_{p}}\left\langle e^{-i\hat{S}}\boldsymbol{\tau}_{a}\cdot\boldsymbol{\tau}_{b}e^{i\hat{S}}\right\rangle _{\textrm{bus}}\right]}{\textrm{Tr}_{p}\left[e^{-\beta H_{p}}\right]}\label{eq:6c}\end{eqnarray}
 It is clear that the canonical transformed Hamiltonian must be a
scalar in the probe's operators, that is, $H_{p}\sim\boldsymbol{\tau}_{a}\cdot\boldsymbol{\tau}_{b}$,
and hence with the same form as found in perturbation theory {[}$H_{\text{eff}}\sim\boldsymbol{\tau}_{a}\cdot\boldsymbol{\tau}_{b}${]}.
However, the operator being averaged with respect to the bus is not
simply $\boldsymbol{\tau}_{a}\cdot\boldsymbol{\tau}_{b}$: \begin{equation}
e^{-i\hat{S}}\boldsymbol{\tau}_{a}\cdot\boldsymbol{\tau}_{b}e^{i\hat{S}}:=\boldsymbol{\tau}_{a}^{R}\cdot\boldsymbol{\tau}_{b}^{R}=\boldsymbol{\tau}_{a}\cdot\boldsymbol{\tau}_{b}-i\left[\hat{S},\boldsymbol{\tau}_{a}\cdot\boldsymbol{\tau}_{b}\right]+....\label{eq:Chap5-Eq7}\end{equation}
 Physical states constructed from the effective Hamiltonian (\ref{eq:effective_hamiltonian})
yield the correct averages for the {}``renormalized'' spins (those
in Fig. \ref{fig:Chap5-CanonicalTransf}, right), but not for the
original spins. Indeed, the spin operators must be renormalized if
one wishes to get averages corresponding to real spin degrees of freedom.
The second term in (\ref{eq:Chap5-Eq7}) gives an important correction
to $\langle\boldsymbol{\tau}_{a}\cdot\boldsymbol{\tau}_{b}\rangle$
{[}Eq.~(\ref{eq:Chap5-Approximation}){]}, inhibiting $\boldsymbol{\tau}_{a(b)}$
partial states with zero linear entropy (in particular, perfect singlets).

Using symmetry alone, we can relate the scalar product involving real
and renormalized spins. The formal derivation is done in Appendix~1
{[}see Eqs.~(\ref{eq:App5-Renormalization-9})-(\ref{eq:tau_tau_can}){]};
here it is sufficient to observe that if the robust gap situation
is verified, taking averages with respect to the spin bus is tantamount
to taking ground-state averages $\langle...\rangle_{\text{bus}}=\langle\psi_{0}|...|\psi_{0}\rangle$;
the result is a probe-only operator that, by symmetry, has the form
\begin{equation}
\langle e^{-i\hat{S}}\boldsymbol{\tau}_{a}\cdot\boldsymbol{\tau}_{b}e^{i\hat{S}}\rangle_{\text{bus}}=\eta\mathbb{I}_{2\otimes2}+(1-\Phi)\boldsymbol{\tau}_{a}\cdot\boldsymbol{\tau}_{b},\label{eq:Chap5-Example1}\end{equation}
 with $\eta=\eta(\alpha)$ and $\Phi=\Phi(\alpha)$, real, bounded,
and temperature independent. No other operators enter in this formula
(\ref{eq:Chap5-Example1}) because the canonical transformation $\hat{S}$
will necessarily produce rotational-invariant probe operators (and
there are just two in $2\otimes2$, namely, the identity and the scalar
product). The parameters $\eta$ and $\Phi$, to which we refer as
canonical corrections, describe how much the states $|\Psi_{0,s}\rangle$
and $|\Psi_{0,m}\rangle$ differ from the product states $|\psi_{0}\rangle\otimes|\chi^{s}\rangle$
and $|\psi_{0}\rangle\otimes|\chi_{m}^{t}\rangle$ {[}see Eqs.~(\ref{eq:Chap5-Eq4})
and (\ref{eq:Chap5-Eq5}){]}.

The probes correlation is obtained by averaging the latter equation.
It is instructive to consider the zero-temperature case,\begin{eqnarray}
\langle\boldsymbol{\tau}_{a}\cdot\boldsymbol{\tau}_{b}\rangle_{T=0} & = & \langle\Psi_{0,s}|\boldsymbol{\tau}_{a}\cdot\boldsymbol{\tau}_{b}|\Psi_{0,s}\rangle\label{eq:Chap5-new1}\\
 & = & \langle\chi^{s},\psi_{0}|\boldsymbol{\tau}_{a}^{R}\cdot\boldsymbol{\tau}_{b}^{R}|\psi_{0},\chi^{s}\rangle\label{eq:Chap5-new2}\\
 & = & -3+\eta+3\Phi.\label{eq:Chap5-new3}\end{eqnarray}
 The last equality implies the restriction $4\ge\eta+3\Phi\ge0$.
The scenario of perfect entanglement, $E(\rho_{ab})=1$, requires
$\eta+3\Phi=0.$ Indeed, considering the approximation of Ref.~\cite{AFF_LDS}
is equivalent to taking $\boldsymbol{\tau}_{a(b)}^{R}\simeq\boldsymbol{\tau}_{a(b)}$,
which results in quasiperfect AF correlations for $T=0$: $\langle\boldsymbol{\tau}_{a}\cdot\boldsymbol{\tau}_{b}\rangle_{T=0}\simeq\langle\chi^{s},\psi_{0}|\boldsymbol{\tau}_{a}\cdot\boldsymbol{\tau}_{b}|\psi_{0},\chi^{s}\rangle=-3.$
This approximation is strictly valid when $\eta\simeq\Phi\simeq0$.
At finite temperatures, we must use the form of the probe Hamiltonian
$H_{p}:=(\Delta_{ab}/4)\boldsymbol{\tau}_{a}\cdot\boldsymbol{\tau}_{b}$,
to obtain, \begin{equation}
\langle\boldsymbol{\tau}_{a}\cdot\boldsymbol{\tau}_{b}\rangle=\eta+(1-\Phi)\langle\boldsymbol{\tau}_{a}\cdot\boldsymbol{\tau}_{b}\rangle_{\textrm{can}}.\label{eq:Chap5-RenormalizedAverage}\end{equation}
 where, \begin{equation}
\langle\boldsymbol{\tau}_{a}\cdot\boldsymbol{\tau}_{b}\rangle_{\textrm{can}}=\frac{\textrm{Tr}_{p}\left[e^{-\beta H_{p}}\boldsymbol{\tau}_{a}\cdot\boldsymbol{\tau}_{b}\right]}{\textrm{Tr}_{p}\left[e^{-\beta H_{p}}\right]}=\frac{e^{-\beta\Delta_{ab}}-1}{e^{-\beta\Delta_{ab}}+1/3}.\label{eq:Chap5-J_can}\end{equation}
 The parameter $\Delta_{ab}$ is, by definition, the singlet-triplet
gap: $\Delta_{ab}=\langle\chi^{t}|H_{p}|\chi^{t}\rangle-\langle\chi^{s}|H_{p}|\chi^{s}\rangle$.
By virtue of the transformation (\ref{eq:Chap5-Eq4}-\ref{eq:Chap5-Eq5})
this equals, \begin{equation}
\Delta_{ab}=\langle\Psi_{0,m}|H|\Psi_{0,m}\rangle-\langle\Psi_{0,s}|H|\Psi_{0,s}\rangle.\label{eq:Chap5-Example4}\end{equation}
 Contrary to the real effective coupling, $J_{ab}(\beta)$, the coupling
$\Delta_{ab}/4$ is effectively temperature independent.

Finally, we relate the real effective coupling defined by Eq.~(\ref{eq:rho_ab})
with the canonical parameters. This is accomplished by equaling the
right-hand sides of Eqs.~(\ref{eq:tau_atau_b}) and (\ref{eq:Chap5-RenormalizedAverage})
and solving for $J_{ab}(\beta)$. We get,\begin{equation}
J_{ab}(\beta)=\frac{1}{4\beta}\ln\left[\frac{3(\Phi-\eta)+(4-3\Phi-\eta)\exp(\beta\Delta_{ab})}{4-\Phi+\eta+(\Phi+\eta/3)\exp(\beta\Delta_{ab})}\right].\label{eq:Jeff as function of Jcan}\end{equation}
We thus have achieved a parametrization of the temperature dependence
of $\rho_{ab}$ as a function of three parameters, the singlet-triplet
gap, $\Delta_{ab},$ and the canonical corrections $\Phi$ and $\eta$:
\begin{equation}
\rho_{ab}(\beta)=\frac{\mathbb{I}_{ab}}{4}+\frac{1}{4}\left(\eta/3+\left(1-\Phi\right)\frac{e^{-\beta\Delta_{ab}}-1}{3e^{-\beta\Delta_{ab}}+1}\right)\boldsymbol{\tau}_{a}\cdot\boldsymbol{\tau}_{b}.\label{eq:Chap5-rho_ab_exacto}\end{equation}
These parameters can in principle be computed in perturbation theory
(see Appendix~2):\begin{eqnarray}
\Delta_{ab} & \simeq & (2J\alpha)^{2}\tilde{\chi}_{r}(0),\label{eq:Chap5-Can1}\\
\Phi & \simeq & (2J\alpha/\sqrt{3})^{2}\sum_{m>0}\sum_{\mu=x,y,z}\frac{|\langle\psi_{0}|\boldsymbol{S}_{A}^{\mu}-\boldsymbol{S}_{B}^{\mu}|\psi_{m}\rangle|^{2}}{\left(E_{m}-E_{0}\right)^{2}},\label{eq:Chap5-Can2}\\
\eta & = & O(\alpha^{4}J^{4}/\Delta_{\textrm{0}}^{4}).\label{eq:Chap5-Can3}\end{eqnarray}
In Eq. (\ref{eq:Chap5-Can2}), the states of the spin bus are denoted
by $|\psi_{m}\rangle$ (with eigenenergy $E_{m}$). The canonical
parameters for small $\alpha$ can be computed by diagonalizing the
spin bus Hamiltonian, $H_{0}|\psi_{k}\rangle=E_{k}|\psi_{k}\rangle$.
This is, however, only possible in a few models whose analytical solutions
are known, as in the case of the 1D XY model \cite{VenutiPRA}, or
in situations where conformal symmetry fixes the form of dynamical
correlations (e.g., critical spin chains) \cite{AFF_LDS}. In general,
whether the canonical parameters describe the correlations of the
probes accurately for a given spin model must be investigated by comparing
result (\ref{eq:Chap5-RenormalizedAverage}) {[}or equivalently, (\ref{eq:Jeff as function of Jcan}){]}
with numerical simulations. We recall that this model will only describe
the partial state of probes interacting with large lattices if adiabatic
continuity and a robust gap hold (Appendix~1). In what follows, we
describe a class of systems that shows impressive agreement with this
canonical theory.%
\begin{figure}
\begin{centering}
\includegraphics[width=0.7\columnwidth]{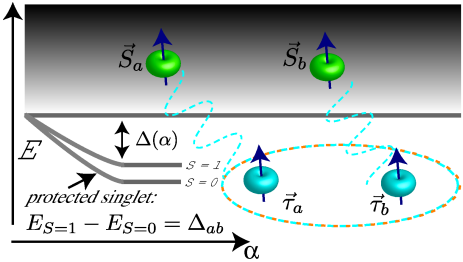} 
\par\end{centering}

\caption{\label{fig:mechanism of LDE}Schematic of the opening of a gap $\Delta_{ab}$
by two spin-$1/2$ probes that couple locally to a bus with arbitrary
strength $\alpha J$. The low-energy sector is separated from excited
states by the gap $\Delta(\alpha)$ and has a singlet-triplet separation,
$\Delta_{ab}$, which is enhanced with the dimensionality of the bus.
If the singlet is localized near the probes, they will be highly entangled
even at large distances (the so-called LDE). We show that $\Delta_{ab}$
is a robust gap in the square AF lattice (see text). }

\end{figure}

\section*{III. Quantum Monte Carlo results}

Our systems consist of 2D finite lattices $\mathcal{L}$, with $N=l\times n_{c}$
spins-$1/2$ and two extra probes, where $l$ is the number of longitudinal
sites and $n_{c}$ stands for the number of coupled chains, varying
from $n_{c}=1$ (spin chain) to $n_{c}=l$ (square lattice) (see Fig.~\ref{fig:Chap5-Lattice}
for a possible geometry). The Hamiltonian of the lattice is \begin{equation}
H_{0}=J\sum_{\langle i,j\rangle}\vec{\boldsymbol{S}}_{i}\cdot\vec{\boldsymbol{S}}_{j},\label{eq:Chap5-LatticeHamiltonian}\end{equation}
with $J>0$. The qubit probes interact with the spins at the boundary
of the most central chain through an isotropic interaction {[}Eq.~(\ref{eq:Chap5-Eq1}){]}.
We expect a significant change in LDE from the common 1D scenario
analyzed in \cite{Venuti,Venuti-b}, as the physics of a 2D spin system
is very distinct. In particular, the 2-leg ladder chain has an Haldane
gap \cite{Book-Auerbach-2} which should play against a large $J_{ab}$;
very \emph{massive excitations}, $\Delta_{0}\simeq0.504J$, make the
correlations die particularly fast \cite{White}. Our QMC simulations
were performed with the library \textsf{looper} from the Algorithms
and Libraries for Physics Simulations (ALPS) project \cite{ALPS}
(see Appendix~3 for details on the numerics).

\subsection*{A.\emph{ }Thermal LDE }

We now outline the main results from the QMC simulations for the entire
family of AF lattices. Among the $20$ spin systems studied, only
the spin chain ($n_{c}=1$) was addressed in previous works. This
system is able to generate a large amount of LDE in the weak coupling
regime, in accordance with the numerical result from \cite{Venuti};
our choice of coupling entails, for $l=20$ and $n_{c}=1$, $J_{p}^{2}/\Delta_{0}\simeq0.015J$
(with $\Delta_{0}$ extracted from \cite{White})---well inside perturbation
limits. The numerical results indeed show probes almost maximally
entangled. Table~\ref{tab:Chap5_CanonicalParameters} below summarizes
the results for the two most representative lattices.

\begin{table}[H]

\noindent \begin{centering}
\begin{tabular}{c||c|c}
\textbf{gaps}  & $\Delta_{0}/J$  & $\Delta_{ab}/J$\tabularnewline
\hline
\hline 
\emph{spin chain}  & $3.2/l\simeq0.16$  & $5.07\times10^{-4}$\tabularnewline
\hline 
\emph{square lattice}  & $9.26/l^{2}\simeq0.02$  & $3.04\times10^{-3}$\tabularnewline
\hline 
\textbf{canonical corrections}  & $\Phi$  & $\eta$\tabularnewline
\hline 
\emph{spin chain}  & $1.03\times10^{-2}$  & $6.27\times10^{-4}$\tabularnewline
\hline 
\emph{square lattice}  & $1.46\times10^{-1}$  & $-1.37\times10^{-2}$\tabularnewline
\end{tabular}
\par\end{centering}

\caption{The singlet-triplet gap $\Delta_{ab}$ and canonical parameters in
representative systems for $\alpha=0.05$. The canonical parameters
are calculated by fitting the QMC data for different temperatures
with Eq.~(\ref{eq:Chap5-RenormalizedAverage}). The expressions for
$\Delta_{\textrm{0}}\equiv\Delta_{ab}(\alpha=0)$ were taken from
Refs. \cite{White} and \cite{Chen}.\label{tab:Chap5_CanonicalParameters}}

\end{table}

These results show a curious property of the 2D spin system: Whereas
for $n_{c}=1$, the spin-triplet gap due to the probes is $\Delta_{ab}\simeq3\times10^{-3}\Delta_{\textrm{0}}$,
for $n_{c}=20$, this gap is $\Delta_{ab}\simeq1.3\times10^{-1}\Delta_{\textrm{0}}$,
a difference of about 2 orders of magnitude for the ratio $\Delta_{ab}/\Delta_{\textrm{0}}$.
This leads to a much more robust LDE against temperature in 2D, albeit
at the cost of a larger canonical correction, $\Phi\sim0.15$, suppressing
the possibility of $E_{C}\simeq1$ at $T=0$. Indeed, taking the values
for $\Phi$ and $\eta$ in table \ref{tab:Chap5_CanonicalParameters}
and using Eqs.~(\ref{eq:Chap5-new3}) and (\ref{eq:Chap5-Concurrence}),
we get, at $T=0$, $E_{C}\simeq0.984$ for the spin chain and $E_{C}\simeq0.788$
for the square lattice. A clear picture of LDE is now given in light
of the exposition of the previous section: The singlet is strongly
localized at the boundary sites (the probes) for $n_{c}=1$, and thus
the mechanism for LDE is quasi optimal, making the effective Hamiltonian
picture {[}Eq.~(\ref{eq:effective_hamiltonian}){]} extremely accurate.
This is not the case in general, where the renormalization of spins
{[}Eq.~(\ref{eq:Chap5-Eq7}){]} cannot be ignored, a fact particularly
evident for the square lattice where $\Phi\sim O(10^{-1})$.

\begin{figure}
\begin{centering}
\includegraphics[width=1\columnwidth]{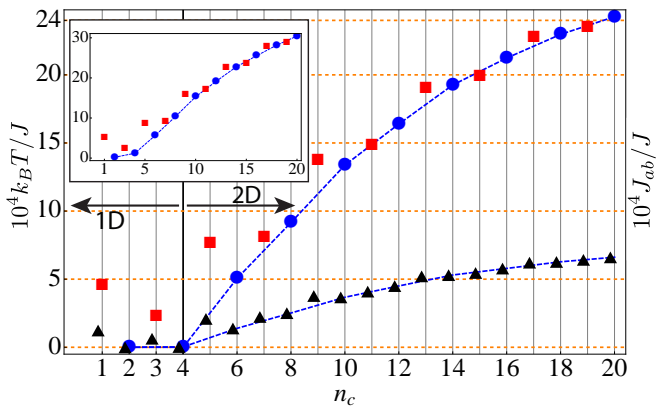} 
\par\end{centering}

\caption{\label{fig:RobustnessGap}Black triangles show effective coupling,
$J_{ab},$ at a distance $l=20$, as a function of $n_{c}$, number
of transverse chains ($k_{B}T=2\times10^{-3}J$ ). Blue dots ($n_{c}$
even) and red squares ($n_{c}$ odd) show temperature above which
LDE vanishes. The inset shows canonical coupling (i.e.,~the probe's
singlet protection gap of Fig.~\ref{fig:mechanism of LDE}) in the
same units used to represent $J_{ab}$. The error bars from QMC cannot
be seen as they are typically below $1$\%; $\alpha=0.05$ in all
plots. }

\end{figure}

\begin{figure}
\noindent \begin{centering}
\includegraphics[width=0.85\columnwidth]{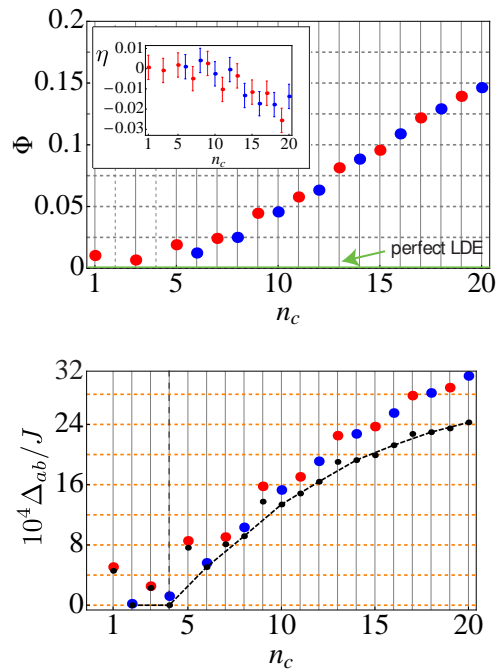} 
\par\end{centering}

\caption{\label{fig:Phi}(top) Canonical correction $\Phi$ for the same family
of spin lattices of Fig.~\ref{fig:RobustnessGap}. The inset shows
that the $\eta$ correction is negligible compared to $\Phi$; the
bars represent an estimate of the error due to QMC fluctuations. The
$n_{c}=2$ and $n_{c}=4$ systems are not represented since the data
do not provide reliable values for canonical corrections. (bottom)
The singlet protection gap for finite probe coupling ($\alpha=0.05$)
in blue ($n_{c}$ even) and red dots ($n_{c}$ odd). The critical
temperature is represented (small black dots) for comparison.}

\end{figure}

Let us investigate these issues more carefully. We postpone the implication
of these results to the emergence of {}``robust gaps'' in 2D. We
focus first on the small probe-bus coupling, $\alpha=0.05$, before
venturing away from perturbation theory. Figure~\ref{fig:RobustnessGap}
shows $J_{ab}$ (for $k_{B}T=2\times10^{-3}J$) and the entanglement
{}``critical temperature'' $T_{c}$. The scale of $T_{c}$ is $\Delta_{ab}$
apart from small corrections (see Fig.~\ref{fig:Phi}, bottom panel).
These plots show a clear enhancement of the ability of the antiferromagnet
to generate long-range effective interactions among distant probes
as one reaches the square lattice at thermal energies $k_{B}T=2\times10^{-3}J$.
The consequences for LDE are evident: The 2D lattices, with $n_{c}>4$,
mediate more entanglement at high temperatures (see later). In Fig.~\ref{fig:RobustnessGap},
a wiggly behavior up to $n_{c}=4$ and a transition for $n_{c}>4$
are also observed; that is, the increase of $\Delta_{ab}$ (and also
$J_{ab}$) becomes smooth because the Haldane gap, very large for
$n_{c}=\left\{ 2,4\right\} $ \cite{White}, gets suppressed.

\begin{figure}
\noindent \begin{centering}
\includegraphics[width=1\columnwidth]{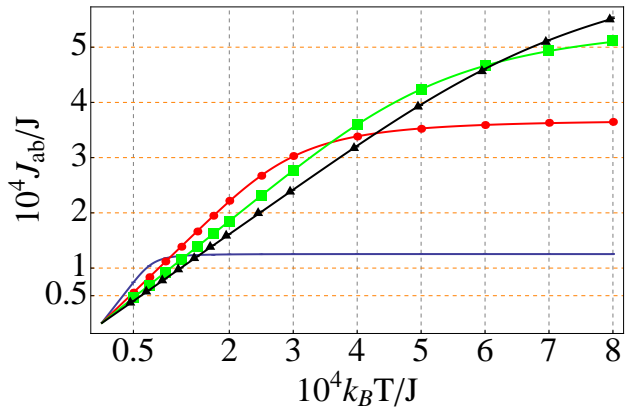} 
\par\end{centering}

\caption{\label{fig:Jeff_Entang_Temp}The points in the plot show $J_{ab}$
as function of the temperature from QMC simulations for $\alpha=0.05$.
The lines stand for the fit with the expression given in Eq.~(\ref{eq:Jeff as function of Jcan}),
with $n_{c}=1$ (blue), $n_{c}=10$ (red; circles), $n_{c}=15$ (green;
squares), and $n_{c}=20$ (black; triangles). The agreement between
the QMC data is excellent, resulting in a average deviation of $\sim0.1-1\%$,
depending on the lattice. }

\end{figure}

We would expect the ground state of $2$D antiferromagnets to reduce
substantially the LDE because of the symmetry breaking at $T=0$,
for large lattices \cite{Review-Manousakis,Chen,Carlson}; the finite
sub-lattice magnetization should reduce the amount of genuine quantum
correlations shared by the probes. This is borne out by the results
of the QMC simulations, shown in Fig.~\ref{fig:Jeff_Entang_Temp},
where $J_{ab}$ (and hence entanglement) is found to decrease at low
temperatures, $k_{B}T\lesssim10^{-4}J$, when the number of chains
increases. Nevertheless, at higher temperatures, the opposite occurs:
$J_{ab}$ increases with $n_{c}$; this reflects the increase of the
probe's protection gap, $\Delta_{ab}$, for it sets the temperature
scale at which entanglement vanishes.

In Fig.~\ref{fig:Jeff_Entang_Temp}, the lines are near perfect fits
to the Monte Carlo data, using Eq.~(\ref{eq:Jeff as function of Jcan}).
For sake of clarity, we have presented the agreement just for four
lattices, although all fits show the same degree of accuracy. The
observed linear dependence of $J_{ab}$ with the temperature for $T\rightarrow0$
is easily understood: A zero-temperature (finite) entanglement below
the maximum value of $1$ requires $\beta J_{ab}(\beta)\rightarrow$constant,
when $\beta\rightarrow\infty$. This constant can be derived from
Eq.~(\ref{eq:Jeff as function of Jcan}), yielding, \begin{equation}
\beta J_{ab}(\beta)\underset{\beta\rightarrow\infty}{\rightarrow}\frac{1}{4}\ln\left[\frac{4-3\Phi-\eta}{\Phi+\eta/3}\right].\label{eq:Chap5-ZeroTemperatureJeff}\end{equation}
Thus, the canonical corrections ($\Phi$ and $\eta$) determine the
low-temperature physics of the probes. It is instructive to notice
that one recovers the condition from quasi-perfect LDE derived earlier
{[}Eq.~(\ref{eq:Chap5-new3}){]}; $\beta J_{ab}\rightarrow\infty$
as $3\Phi+\eta\rightarrow0$ and thus, according to Eq.~(\ref{eq:tau_atau_b}),
$\langle\boldsymbol{\tau}_{a}\cdot\boldsymbol{\tau}_{b}\rangle_{T=0}\simeq-3$.
This is a very special scenario occurring in 1D antiferromagnets and
also in dimerized chains \cite{VenutiPRA}. The present results reveal
that quasiperfect LDE is also mediated by the three-ladder chain,
$3\Phi+\eta\simeq0.007$. As soon as we approach the 2D scenario,
more precisely when $n_{c}>4$, the $\Phi$ correction gets larger
(see Fig.~\ref{fig:Phi} for the variation of $\Phi$ with $n_{c}$)
and a fraction of the entanglement is lost.

In our simulations, the value of $\eta$ is negligible (a careful
inspection shows that the fits we present are virtually indistinguishable
from the fits with $\eta=0$ up to $\alpha\simeq0.1$), and the $n_{c}=3$
spin system mediates the largest amount of LDE at $T=0$ (see Fig.~\ref{fig:Phi}),
$E_{C}\simeq0.997$. Curiously, this behavior is not altered by varying
the coupling in the entire range we have simulated: $\alpha\in[0.05,0.2]$.
In fact, for large $\alpha$, namely, $\alpha=0.2$, the discrepancy
between the $n_{c}=1$ and $n_{c}=3$ lattices is quite significant:
$\Phi\simeq0.238$ against $\Phi\simeq0.079$. This is a very interesting
property of the three-leg ladder: the ability of generating high-quality
ground-state entanglement for a wide range of probe-bus coupling.

For the highest temperatures simulated, the effective coupling saturates
(Fig.~\ref{fig:Jeff_Entang_Temp}) to a constant value, $\Delta_{ab}(1-\Phi)/4$,
when $\eta$ is negligible (i.e\emph{.},~not far away from the perturbation
limit, $\alpha\lesssim0.1$), suffering a slight change with temperature
otherwise, \begin{equation}
J_{ab}(T)\simeq\frac{\Delta_{ab}}{4}\left(1-\Phi\right)-\frac{k_{B}T}{12}\eta+O(\frac{\Delta_{ab}^{2}\Phi}{k_{B}T}).\label{eq:Jeff-high_temp_expan}\end{equation}
An estimate of the entanglement critical temperature, $T_{c}$, is
obtained by noting that $J_{ab}(T)$ has already saturated when the
concurrence vanishes. Indeed, combining Eqs.~(\ref{eq:critical temperature})
and (\ref{eq:Jeff-high_temp_expan}), we obtain,\begin{equation}
k_{B}T_{c}\simeq0.93\Delta_{ab}(1-\Phi).\label{eq:Chap5-CriticalTemp}\end{equation}
This agrees with the numerical results within $1\%$; in Fig.~\ref{fig:Phi}
(bottom), the increasing mismatch between $k_{B}T_{c}$ and the gap
$\Delta_{ab}$ with $n_{c}$ reflects the increase in $\Phi$. This
equation for $T_{c}$ generalizes the previous result for the spin-$1/2$
AF Heisenberg ring \cite{AFF_LDS}---where $\Delta_{ab}$ is equal
to $4\alpha^{2}\tilde{\chi}_{r}(0)$ in the perturbative regime {[}Eq.~(\ref{eq:Chap5-Can1}){]}---by
the inclusion of the $\Phi$ correction.

The square lattice is the system with the best thermal LDE robustness,
despite its appreciable reduction of zero temperature entanglement,
$\Phi\simeq0.146$, a fact explained by the emergence of a large singlet-triplet
gap, $\Delta_{ab}$, which is about $6$ times the protection gap
of the single spin chain. The regime of high temperatures, $k_{B}T\gtrsim\Delta_{0}$,
is not described by Eq.~(\ref{eq:Jeff as function of Jcan}) anymore,
which assumes negligible thermal occupancy of excited states of the
spin bus, a crucial assumption of our analytical modeling. However,
according to our estimate {[}Eq.~(\ref{eq:Chap5-CriticalTemp}){]},
no LDE is expected in this temperature range, because it vanishes
at much lower temperatures; $J_{ab}(\beta)$ should decrease with
the temperature at some point and eventually drop to zero, reflecting
totally uncorrelated probes. The initial drop of $J_{ab}$ with $T$
is captured by Eq.~(\ref{eq:Jeff-high_temp_expan}).

\subsection*{B. Beyond weak coupling}

\begin{figure}
\noindent \begin{centering}
\includegraphics[width=0.7\columnwidth]{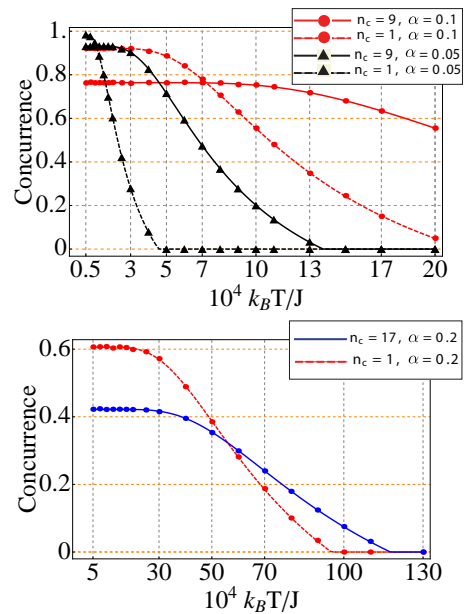} 
\par\end{centering}

\caption{\label{fig:Entang_Various}Concurrence, an entanglement monotone for
qubits, as a function of temperature for different lattices and couplings.
Lattices supporting more entanglement at strictly $T=0$ have worse
performance at higher temperatures. The lines are fits using Eq.~(\ref{eq:Jeff as function of Jcan})
for $J_{ab}(\beta)$ and Eqs.~(\ref{eq:tau_atau_b}) and (\ref{eq:Chap5-Concurrence}).
The fits are nearly perfect for the entire temperature range in which
entanglement persists.}

\end{figure}

Figures~\ref{fig:RobustnessGap} and \ref{fig:Phi} deal with relatively
small probe-bus coupling, but the results presented so far are more
general. For instance, choosing a sufficiently large $\alpha$ to
strongly suppress the zero-temperature entanglement via partial frustration
among the neighborhood of the bulk spins connected with the probes,
we again find excellent agreement with Eq.~(\ref{eq:Jeff as function of Jcan}).
For intermediate probe-bus coupling, $\alpha=0.1$ and $\alpha=0.2$,
the measured concurrence is fitted with an expression derived from
Eq.~(\ref{eq:Chap5-RenormalizedAverage}), as shown in Fig.~\ref{fig:Entang_Various}.
These results show that in all our measured systems, the conditions
for our model hold, namely, a singlet-triplet low-energy sector and
an excited bus sector which is not significantly populated. This is
surprising, particularly in the case of large lattices, which have
a gap $\Delta_{0}$ smaller than $\alpha J$; not only do we find
a well protected singlet with a considerable amount of LDE, but the
lowest singlet and triplet remain separated from the rest of the spectrum,
allowing a complete description of entanglement only in terms of these
two energy levels.

On the other hand, though the strong coupling to the spin bus reduces
the zero temperature entanglement, it also allows a larger split between
the singlet and triplet, leading to entangled probes at much higher
temperatures. Typically, exchange interactions in antiferromagnets
can be of the order of $0.1\mathrm{\, eV}$, resulting in an effective
coupling of the order of $0.3\,\mathrm{meV}$ for the square lattice
($l=20)$ at temperature $\sim12\,\mathrm{K}$ and $\alpha=0.2$.
This is to be compared with the value of $0.01-0.1\mathrm{\, me}\mathrm{V}$
achievable in quantum dot spins \cite{DiVincenzo3}, although decoherence
effects in spin lattices can lessen this difference. The critical
temperature (above which the correlations shared by the probes are
completely classical) can be increased by a factor of $20$ from a
weakly coupled spin chain ($\alpha=0.05)$ to an intermediate coupled
($\alpha=0.2)$ 2D lattice, entanglement surviving up to $k_{B}T\simeq1.2\times10^{-2}J$
(Fig.~\ref{fig:Entang_Various})---an appreciable enhancement of
the thermal robustness of such correlations.

\subsection*{C. Robust gaps in 2D}

The QMC results analyzed in light of our model show that two probes
interacting with AF spin systems open a gap $\Delta_{ab}$ which is
enhanced with the dimensionality of the system. For the square lattice
and probe-bus coupling $J_{p}=0.05J$, the gap reads $\Delta_{ab}\simeq3.04\times10^{-3}J$
, a \emph{robust gap}. The point is that the $20\times20$ lattice,
without probes, has a small singlet-triplet gap $\Delta_{0}\simeq0.02J$
(see Refs.~\cite{Chen,Carlson}), making the bus-probe interaction
non-perturbative. Even so, these probes opened a considerable gap,
and their reduced state displays a large amount of $T=0$ entanglement,
$E_{C}\simeq0.79$, in resemblance to the spin chain bus with probes.
The situation in the 2D lattice is very distinct, though, because
for the spin chain, $\Delta_{ab}/\Delta_{\textrm{0}}=O(10^{-3})$,
whereas in the square lattice, this ratio is $2$ orders of magnitude
greater, $\Delta_{ab}/\Delta_{\textrm{0}}=O(10^{-1})$.

\begin{figure}
\begin{centering}
\includegraphics[width=0.9\columnwidth]{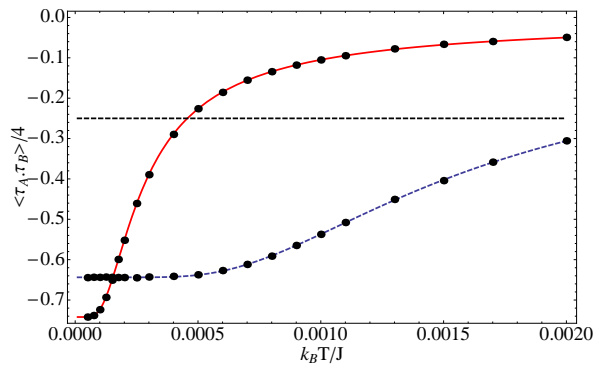} 
\par\end{centering}

\caption{\label{fig:Probe-probe-correlation}Probe-probe correlation, $\langle\boldsymbol{\tau}_{a}\cdot\boldsymbol{\tau}_{b}\rangle/4$,
at a distance $l=20$, as a function of temperature for the spin chain
(red) and the square lattice (blue); $\alpha=0.05$ in both plots.
The black dots are given by the QMC simulations (the error bars from
QMC cannot be seen as they are typically below $1$\%). The lines
stand for the fit with Eq. (\ref{eq:Chap5-RenormalizedAverage}).
The dashed line separates classical correlations (above) from quantum
correlations (below). The red curve shows that the quality of the
fits extends beyond the region where entanglement is nonzero.}

\end{figure}

The whole analysis of this article follows from fits of the probe-probe
correlation $\langle\boldsymbol{\tau}_{a}\cdot\boldsymbol{\tau}_{b}\rangle$
computed by QMC for several temperatures to a simple analytical model
derived under adiabatic continuity hypothesis. These fits provide
the values of $\Delta_{ab}$ and the canonical corrections ($\Phi$
and $\eta$). Our reliance on this model is not exclusively due to
the nearly perfect fits to which it leads (see Fig.~\ref{fig:Probe-probe-correlation}).

\begin{figure}
\centering{}\includegraphics[width=0.9\columnwidth]{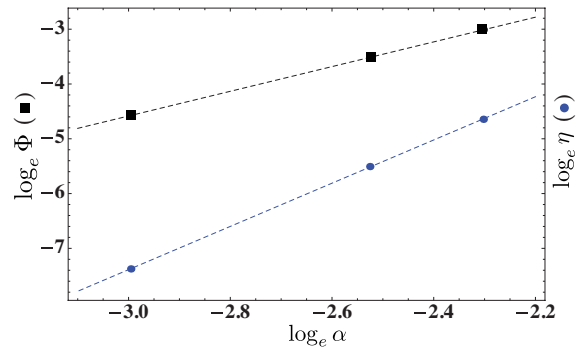}\caption{\label{fig:FIT}Log-log representation (dots) of $\Phi(\alpha)$ and
$\eta(\alpha)$ for a spin chain with $L=20$ plus two probes---the
$n_{c}=1$ system of the article---and the respective linear fits
(dashed lines). }

\end{figure}

A crucial test of this model is given by checking the scaling of $\Phi$
and $\eta$ with $\alpha$ {[}Eqs.~(\ref{eq:Chap5-Can2}) and (\ref{eq:Chap5-Can3}){]}.
We give an estimate for this scaling for the spin chain with $L=20$
based on three simulations in Fig.~\ref{fig:FIT}: The dependence
of $\Phi$ and $\eta$ with $\alpha$ can be read from the slopes;
$\Phi\sim\alpha^{2.26}$ and $\eta\sim\alpha^{3.95}$, which agrees
well with the theoretical prediction for weak coupling given by our
model ($\Phi\sim\alpha^{2}$ and $\eta\sim\alpha^{4}$). Recall that
for $\alpha=0.2$, we have $\alpha J>\Delta_{\textrm{0}}$, and thus
the system is already well inside the intermediate coupling regime.
Nevertheless, the fits are very good up this value of coupling (see
Fig.~\ref{fig:Entang_Various})---we again emphasize that this is
particularly remarkable for the 2D system, which is gapless in the
thermodynamic limit and has a broken symmetry phase. These facts provide
added confirmation of our analytical model and the conclusions drawn
about the spectrum of the system and the emergence of robust gaps.%

\begin{figure}
\begin{centering}
\includegraphics[width=0.5\columnwidth]{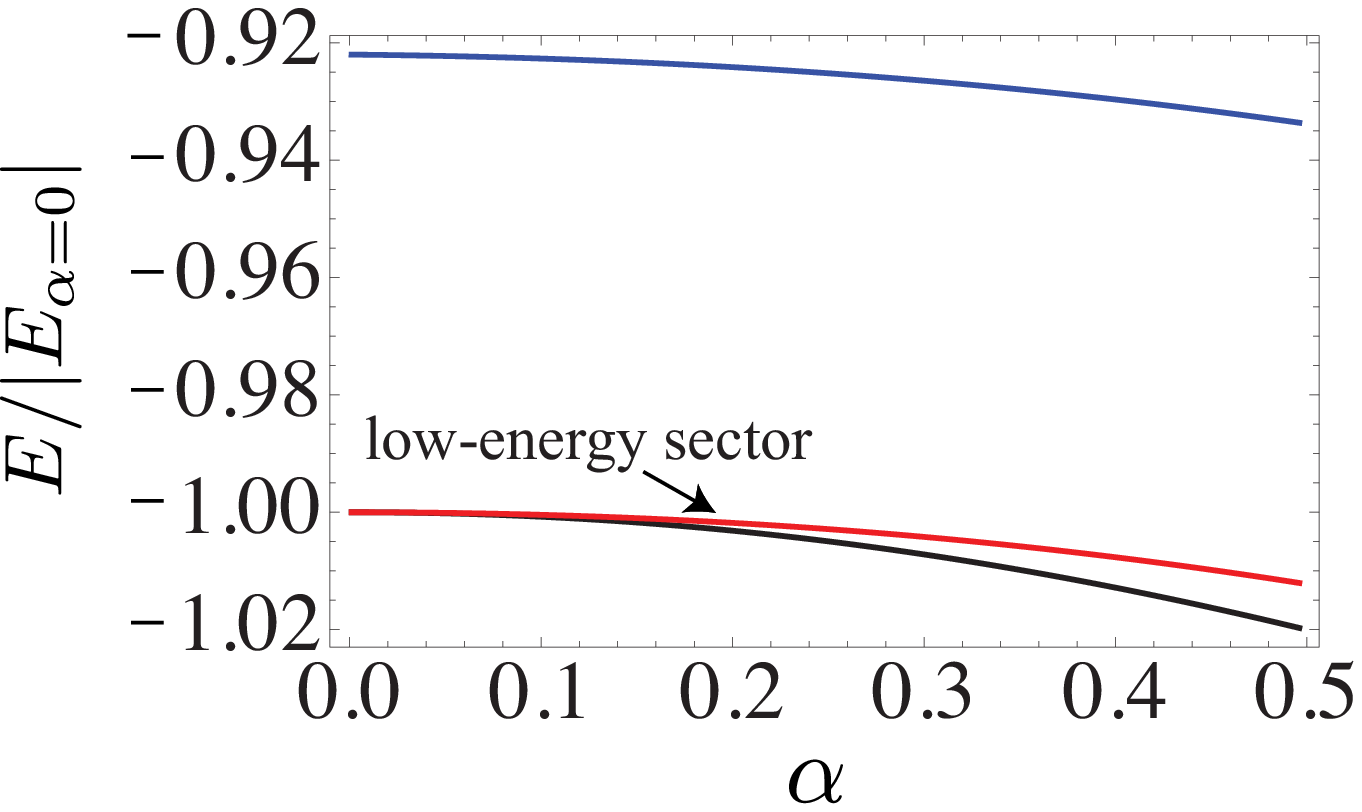}\includegraphics[width=0.5\columnwidth]{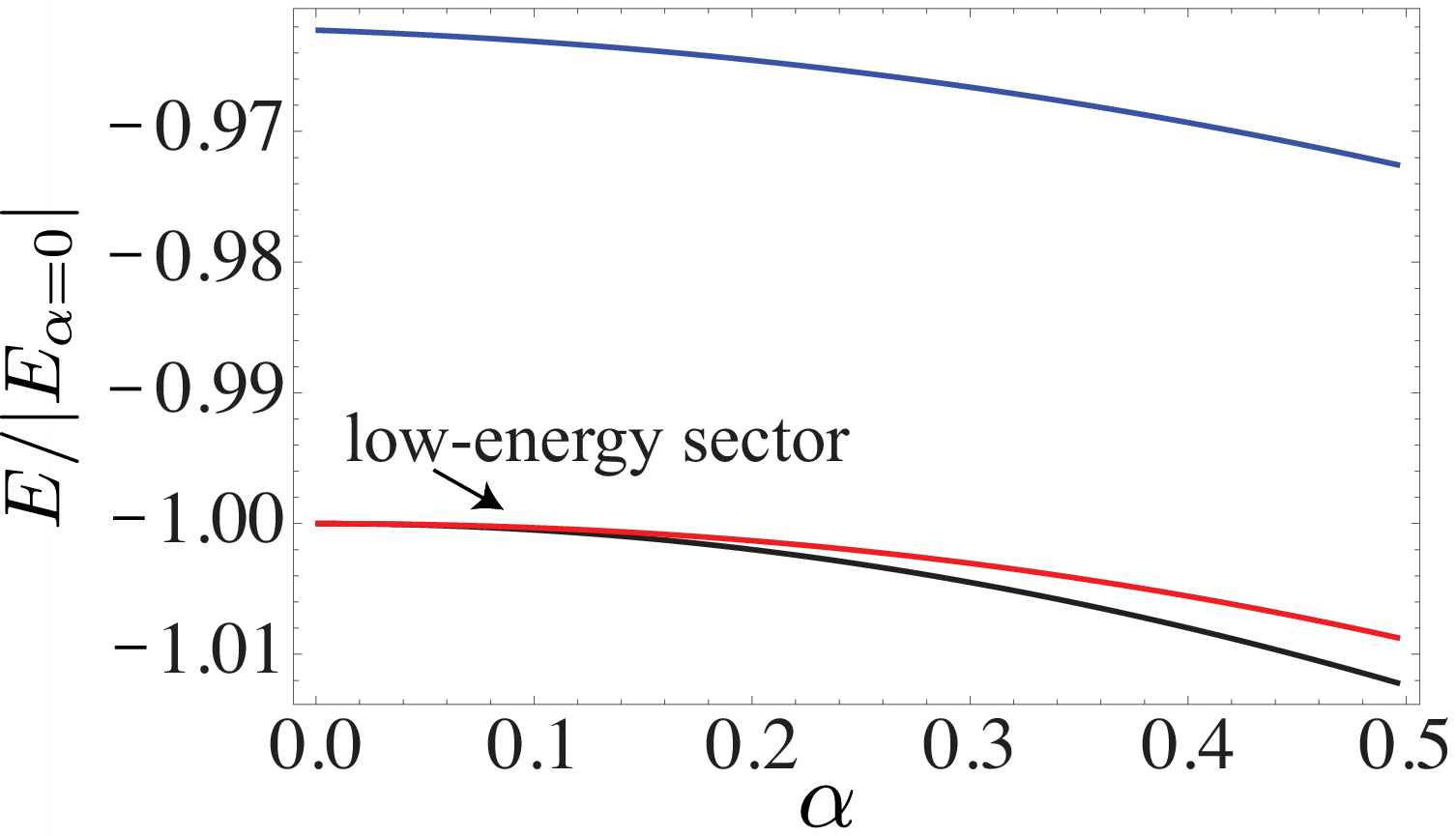} 
\par\end{centering}

\caption{\label{fig:SpectrumSmall}Plot of the low-energy spectrum for two
different systems with $100$ values of $\alpha$ in the range $[0,0.5]$.
The energy is scaled to the absolute value of the ground-state energy
for $\alpha=0$: (left) a lattice with $l=4$ and $n_{c}=3$ with
gap $\Delta_{0}=0.52J$ and (right) a lattice with $l=6$ and $n_{c}=3$
with gap $\Delta_{0}=0.39J$. The crossover between weak and strong
probe-bus coupling can be observed with the respective emergence of
robust gaps.}

\end{figure}

If states other than the lowest singlet and triplet contributed significantly
to the spin correlations at the temperatures in which we still find
entanglement, the quality of the fits, added to the fact that $\eta\ll\Phi$,
would be an amazing coincidence indeed. Exact numerical diagonalizations
could provide direct evidence of the spectrum, but these are very
hard to perform for large 2D systems. We have used the algorithm \textsf{sparsediag}
from ALPS to compute the three lowest energies for relatively small
systems; Fig.~\ref{fig:SpectrumSmall} shows the emergence of robust
gaps for two different lattices ($4\times3$ and $6\times3)$. %

\section*{IV. Conclusions}

We observed an enhancement of LDE by varying the geometry of the magnetic
spin systems serving as a quantum bus. We presented the first \emph{direct}
measurement of the probe correlation for finite temperature by means
of QMC simulations with the ALPS algorithm. Also, we derived an analytical
expression for the probe density matrix that completely describes
the probe-probe entanglement for the systems we studied, even away
from the weak-coupling regime. The importance of the dimensionality
of the bus is twofold: On one hand, ground-state entanglement is reduced
as we go from a 1D to a 2D bus; on the other hand, robustness of thermal
entanglement increases considerably in the 2D, raising the possibility
of entangling distant spin probes at temperatures as high as $T\simeq1.2\times10^{-2}J/k_{B}$,
where $J$ is the nearest neighbor exchange constant of the bus, an
appealing situation for quantum computation based on magnetic spins.
Finally, we reported on a spin system supporting quasiperfect ground-state
LDE more efficiently than the AF spin chain, particularly in the intermediate
probe-bus coupling regime: the three-leg ladder chain. 
\begin{acknowledgments}
We acknowledge the ALPS project \cite{ALPS} and the library \textsf{looper}
that made possible the QMC simulations of the present article, and
we also acknowledge the support of EU and FCT through POCI 2010 and
PRAXIS Grant No. SFRH/BD/18292/04 (A.F.) and Grant No. SFRH/BPD/27160/2006
(J.V.L.). 
\end{acknowledgments}
\appendix

\section{Appendix 1}

Full rotational symmetry entails that the probe density matrix can
be written as function of a single $2\otimes2$ invariant,\begin{equation}
\hat{\rho}_{ab}=\frac{1}{3e^{-\beta J_{ab}(\beta)}+e^{3J_{ab}(\beta)}}\exp\left(-\beta J_{ab}(\beta)\boldsymbol{\tau}_{a}\cdot\boldsymbol{\tau}_{b}\right),\label{eq: App5-eq:rho_ab}\end{equation}
 This allows to parametrize the correlations between probes as \begin{equation}
\langle\boldsymbol{\tau}_{a}\cdot\boldsymbol{\tau}_{b}\rangle=\frac{e^{-4\beta J_{ab}(\beta)}-1}{e^{-4\beta J_{ab}(\beta)}+1/3}.\label{Eq: App5-eq:tau_atau_b}\end{equation}
 This tells us very little for the moment since the temperature dependence
of $J_{ab}(\beta)$ is unknown. On the other hand, by definition,\begin{equation}
\langle\boldsymbol{\tau}_{a}\cdot\boldsymbol{\tau}_{b}\rangle=\frac{\textrm{Tr}\left[e^{-\beta H}\boldsymbol{\tau}_{a}\cdot\boldsymbol{\tau}_{b}\right]}{\textrm{Tr}\left[e^{-\beta H}\right]},\label{eq:App5-Renormalization-8}\end{equation}
 and, using the canonical transformation $\hat{S}$, \begin{equation}
\langle\boldsymbol{\tau}_{a}\cdot\boldsymbol{\tau}_{b}\rangle=\frac{\textrm{Tr}\left[e^{-\beta H_{S}}e^{-i\hat{S}}\boldsymbol{\tau}_{a}\cdot\boldsymbol{\tau}_{b}e^{i\hat{S}}\right]}{\textrm{Tr}\left[e^{-\beta H_{S}}\right]}.\label{eq:App5-Renormalization-9}\end{equation}
 The transformed Hamiltonian is $H_{S}=H_{p}+H_{b}^{'}$; the corresponding
eigenbasis is made of product states {[}Eqs.~(\ref{eq:Chap5-Eq4})-(\ref{eq:Chap5-Eq5}){]}.
Under the assumption that $k_{B}T\ll\Delta(\alpha)$ (i.e., that the
temperature is much smaller than the gap to excited states of the
bus), we can limit the trace to the states of the form $|\psi_{0}\rangle\otimes|\chi\rangle$,
where $|\chi\rangle$is any probe state and $|\psi_{0}\rangle$ is
the nondegenerate ground state of the spin bus. This leads to \begin{equation}
\langle\boldsymbol{\tau}_{a}\cdot\boldsymbol{\tau}_{b}\rangle=\frac{\textrm{Tr}_{p}\left[e^{-\beta H_{p}}\langle\psi_{0}|e^{-i\hat{S}}\boldsymbol{\tau}_{a}\cdot\boldsymbol{\tau}_{b}e^{i\hat{S}}|\psi_{0}\rangle\right]}{\textrm{Tr}_{p}\left[e^{-\beta H_{p}}\right]},\label{eq:tau_atau_n_canon}\end{equation}
 where $\textrm{Tr}_{p}(\dots)$ is a trace over probe states only.
Since the operator $\boldsymbol{\tau}_{a}\cdot\boldsymbol{\tau}_{b}$
is diagonal in bus space, this can obviously be written as\begin{eqnarray}
\langle\boldsymbol{\tau}_{a}\cdot\boldsymbol{\tau}_{b}\rangle & =\frac{1}{\textrm{Tr}_{p}\left[e^{-\beta H_{p}}\right]}\textrm{Tr}_{p} & \left[e^{-\beta H_{p}}\sum_{m}\langle\psi_{0}|e^{-i\hat{S}}|\psi_{m}\rangle\right.\nonumber \\
 &  & \left.\boldsymbol{\tau}_{a}\cdot\boldsymbol{\tau}_{b}\langle\psi_{m}|e^{i\hat{S}}|\psi_{0}\rangle\right]\label{eq:App5-Renormalization-10}\end{eqnarray}
 \begin{equation}
=\frac{\textrm{Tr}_{p}\left[e^{-\beta H_{p}}\sum_{m}A_{m}\boldsymbol{\tau}_{a}\cdot\boldsymbol{\tau}_{b}A_{m}^{\dagger}\right]}{\textrm{Tr}_{p}\left[e^{-\beta H_{p}}\right]},\label{eq:tau_atau_b_am}\end{equation}
 where $A_{m}\equiv\langle\psi_{0}|e^{-i\hat{S}}|\phi_{m}\rangle$
is an operator defined in probe space. By symmetry, the operator \begin{equation}
\sum_{m}A_{m}\boldsymbol{\tau}_{a}\cdot\boldsymbol{\tau}_{b}A_{m}^{\dagger}=\langle\psi_{0}|e^{-i\hat{S}}\boldsymbol{\tau}_{a}\cdot\boldsymbol{\tau}_{b}e^{i\hat{S}}|\psi_{0}\rangle\label{eq:App5-DefinitionAm}\end{equation}
 must be a scalar in probe space and therefore of the form \begin{equation}
\sum_{m}A_{m}\boldsymbol{\tau}_{a}\cdot\boldsymbol{\tau}_{b}A_{m}^{\dagger}=\eta+\left(1-\Phi\right)\boldsymbol{\tau}_{a}\cdot\boldsymbol{\tau}_{b},\label{eq:eta_phi}\end{equation}
 where $\eta$ and $\Phi$ are, by construction, temperature-independent
renormalization constants. Since $\textrm{Tr}_{p}\left[\boldsymbol{\tau}_{a}\cdot\boldsymbol{\tau}_{b}\right]=0$,
and \begin{equation}
(\boldsymbol{\tau}_{a}\cdot\boldsymbol{\tau}_{b})^{2}=3-2\boldsymbol{\tau}_{a}\cdot\boldsymbol{\tau}_{b},\label{eq:App5-Renormalization-11}\end{equation}
 we obtain\begin{eqnarray}
\textrm{Tr}_{p}\left[\sum_{m}A_{m}\boldsymbol{\tau}_{a}\cdot\boldsymbol{\tau}_{b}A_{m}^{\dagger}\right] & = & 4\eta,\label{eq:eta:def}\\
\textrm{Tr}_{p}\left[\sum_{m}\boldsymbol{\tau}_{a}\cdot\boldsymbol{\tau}_{b}A_{m}\boldsymbol{\tau}_{a}\cdot\boldsymbol{\tau}_{b}A_{m}^{\dagger}\right] & = & 12(1-\Phi).\label{eq:phi_def}\end{eqnarray}
 With these definitions it is clear that \begin{equation}
\langle\boldsymbol{\tau}_{a}\cdot\boldsymbol{\tau}_{b}\rangle=\eta+(1-\Phi)\langle\boldsymbol{\tau}_{a}\cdot\boldsymbol{\tau}_{b}\rangle_{\mathrm{\textrm{can}}},\label{eq:tau_tau_parameters}\end{equation}
 where,\begin{equation}
\langle\boldsymbol{\tau}_{a}\cdot\boldsymbol{\tau}_{b}\rangle_{\textrm{can}}:=\frac{\textrm{Tr}_{p}\left[e^{-\beta H_{p}}\boldsymbol{\tau}_{a}\cdot\boldsymbol{\tau}_{b}\right]}{\textrm{Tr}_{p}\left[e^{-\beta H_{p}}\right]}=\frac{e^{-\beta\Delta_{ab}}-1}{e^{-\beta\Delta_{ab}}+1/3}.\label{eq:tau_tau_can}\end{equation}
This looks exactly like the preceding expression, except that now
$\Delta_{ab}$, unlike $J_{ab}$, is temperature independent.\emph{
}So we achieve a parametrization of $\langle\boldsymbol{\tau}_{a}\cdot\boldsymbol{\tau}_{b}\rangle$
in terms of three temperature-independent parameters $\Delta_{ab},$
$\Phi$, and $\eta$. This result, although simple, has important
consequences; for instance, we see that symmetry implies that $\Delta_{ab}$
is in fact the gap separating the probe's singlet and the probe's
triplet up to any order. We can thus write,\begin{equation}
\Delta_{ab}=E_{\textrm{triplet}}(\alpha)-E_{\textrm{singlet}}(\alpha).\label{eq:App5-Renormalization-ExactGap}\end{equation}
Using equations~(\ref{Eq: App5-eq:tau_atau_b}), (\ref{eq:tau_tau_parameters})
and (\ref{eq:tau_tau_can}), we can express $J_{ab}(\beta)$ explicitly
in these temperature-independent parameters:\begin{equation}
J_{ab}(\beta)=\frac{1}{4\beta}\ln\left[\frac{3(\Phi-\eta)+(4-3\Phi-\eta)\exp(\beta\Delta_{ab})}{4-\Phi+\eta+(\Phi+\eta/3)\exp(\beta\Delta_{ab})}\right].\label{eq:App5-Renormalization-13}\end{equation}
In Appendix 2 we derive the expression for $\Phi$, and show that
it is of second order in the small parameter $J\alpha/\Delta$.

\section{Appendix 2}

We now compute the canonical corrections, $\Phi$ and $\eta$, to
the probe canonical correlation {[}Eq.~(\ref{eq:tau_tau_can}){]}
in perturbation theory. We assume the following conditions to hold:
(1) the temperature is small enough not to generate real excitations
of the spin lattice system and (2) the probes couple weakly to the
spin bus via an isotropic interaction with strength $\alpha J$, such
that $\alpha J\ll\Delta_{0}$ . In this limit, we can use the Schrieffer-Wolff
prescription for the canonical generator $\hat{S}$ \cite{Wolff},
\[
V+i\left[H_{0},\hat{S}\right]=0,\]
where $H_{0}$ is the bus Hamiltonian and $V$ the bus-probe coupling.
The canonical generator $\hat{S}$ has nonzero matrix elements only
between the ground state manifold (at zero coupling) and excited states
of order $\sim\mathcal{O}(\alpha J/\Delta_{0}).$ The transformed
Hamiltonian is, to second order in $V$, \[
H_{S}:=H_{0}-\frac{i}{2}\left[\hat{S},V\right].\]
To calculate the probe correlation, we expand the renormalized spin
operators in powers of $\hat{S}:$

\begin{eqnarray}
\langle\boldsymbol{\tau}_{a}\cdot\boldsymbol{\tau}_{b}\rangle & = & \textrm{Tr}\left[e^{-i\hat{S}}\rho_{ab}e^{i\hat{S}}e^{-i\hat{S}}\boldsymbol{\tau}_{a}\cdot\boldsymbol{\tau}_{b}e^{i\hat{S}}\right]\label{eq:App5-Canonical-1}\\
 & = & \textrm{Tr}\left[\frac{e^{-\beta H_{S}}}{\mathcal{Z}}\left(\boldsymbol{\tau}_{a}\cdot\boldsymbol{\tau}_{b}-i\left[\hat{S},\boldsymbol{\tau}_{a}\cdot\boldsymbol{\tau}_{b}\right]\right.\right.\nonumber \\
 &  & \left.\left.-\frac{1}{2}\left[\hat{S},\left[\hat{S},\boldsymbol{\tau}_{a}\cdot\boldsymbol{\tau}_{b}\right]\right]+O(\frac{J^{3}\alpha^{3}}{\Delta_{0}^{3}})\right)\right],\label{eq:App5-Canonical-3}\end{eqnarray}
with $\mathcal{Z}:=\text{Tr}\left[\exp\left(-\beta H_{S}\right)\right]$.
From the latter result, we can express the $\hat{S}$-renormalized
spins as a function of the original spins: \begin{equation}
\boldsymbol{\tau}_{a}\cdot\boldsymbol{\tau}_{b}\underset{\hat{S}}{\longrightarrow}\boldsymbol{\tau}_{a}^{R}\cdot\boldsymbol{\tau}_{b}^{R}=\boldsymbol{\tau}_{a}\cdot\boldsymbol{\tau}_{b}+\boldsymbol{\xi}_{ab},\label{eq:App5-Canonical-3b}\end{equation}
 with,\begin{equation}
\boldsymbol{\xi}_{ab}=-\frac{1}{2}\left[\hat{S},\left[\hat{S},\boldsymbol{\tau}_{a}\cdot\boldsymbol{\tau}_{b}\right]\right]+O(J^{3}\alpha^{3}\Delta_{0}^{-3}),\label{eq:Correction to Probes Operators}\end{equation}
implying that the ground state of the probes $\rho_{ab}$ will never
be a perfect singlet if $\langle\boldsymbol{\xi}_{ab}\rangle_{\beta}$
is nonnegligible. The trace in Eq.~(\ref{eq:App5-Canonical-3}) can
be performed in two steps: (1) tracing out the bus by considering
just the overlap with the ground-state, which is justified by the
low-temperature condition, and (2) performing a thermal average in
the $2\otimes2$ Hilbert space of the probes.

\noindent The first-order term does not contribute as the generator
$\hat{S}$ has null matrix elements in the ground state, \emph{\begin{equation}
\langle\psi_{0}|\left[\hat{S},\boldsymbol{\tau}_{a}\cdot\boldsymbol{\tau}_{b}\right]|\psi_{0}\rangle=\left[\langle\psi_{0}|\hat{S}|\psi_{0}\rangle,\boldsymbol{\tau}_{a}\cdot\boldsymbol{\tau}_{b}\right]=0.\label{eq:App5-Canonical-4}\end{equation}
}Then we are left with a zero-order term,\[
C_{0}=\text{Tr}\left[\frac{e^{-\beta H_{S}}}{\mathcal{Z}}\boldsymbol{\tau}_{a}\cdot\boldsymbol{\tau}_{b}\right],\]
 yielding the canonical correlation, $\langle\tau_{a}\cdot\tau_{b}\rangle_{\textrm{can}}$
, and with a second-order correction $C_{2}$,\begin{equation}
C_{2}=-\frac{1}{2\mathcal{Z}}\textrm{Tr}\left\{ e^{-\beta H_{S}}\left[\hat{S},\left[\hat{S},\boldsymbol{\tau}_{a}\cdot\boldsymbol{\tau}_{b}\right]\right]\right\} .\label{eq:Correction_GeneratorForm}\end{equation}
We must have some care to evaluate the preceding thermal average.
Let us reproduce the main steps,\begin{eqnarray}
C_{2} & = & -\frac{1}{2\mathcal{Z}}\textrm{Tr}\left[e^{-\beta H_{S}}\left(\hat{S}\hat{S}\boldsymbol{\tau}_{a}\cdot\boldsymbol{\tau}_{b}\right.\right.\nonumber \\
 &  & \left.\left.-2\hat{S}\boldsymbol{\tau}_{a}\cdot\boldsymbol{\tau}_{b}\hat{S}+\boldsymbol{\tau}_{a}\cdot\boldsymbol{\tau}_{b}\hat{S}\hat{S}\right)\right]\label{eq:App5-Canonical-5}\\
 & \simeq & -\frac{1}{\mathcal{Z}}\sum_{k\ge0}\textrm{Tr}_{a,b}\left[\langle\psi_{0}|e^{-\beta H_{S}}|\psi_{k}\rangle\right.\nonumber \\
 &  & \left.\langle\psi_{k}|\left(\hat{S}\hat{S}\boldsymbol{\tau}_{a}\cdot\boldsymbol{\tau}_{b}-\hat{S}\boldsymbol{\tau}_{a}\cdot\boldsymbol{\tau}_{b}\hat{S}\right)|\psi_{0}\rangle\right].\label{eq:App5-Canonical-7}\end{eqnarray}
The effective Hamiltonian has no matrix elements between eigenstates
belonging to different sectors (up to the order at which we are working),
which simplifies the above summation as only the ground-state contributes
{[}compare with Eqs.~(\ref{eq:App5-Renormalization-9})-(\ref{eq:tau_atau_n_canon}){]}:
\begin{equation}
\langle\psi_{0}|e^{-\beta H_{S}}|\psi_{k}\rangle=e^{-\beta H_{p}}\langle\psi_{0}|e^{-\beta H_{b}^{\prime}}|\psi_{k}\rangle\delta_{0k}.\label{eq:App5-Canonical-7b}\end{equation}
Indeed,\begin{equation}
C_{2}=-\frac{1}{\mathcal{Z}}\text{Tr}_{a,b}\left\{ e^{-\beta H_{p}}\langle\psi_{0}|\left(\hat{S}\hat{S}\boldsymbol{\tau}_{a}\cdot\boldsymbol{\tau}_{b}-\hat{S}\boldsymbol{\tau}_{a}\cdot\boldsymbol{\tau}_{b}\hat{S}\right)|\psi_{0}\rangle\right\} \label{eq:App5-Canonical-10}\end{equation}

\noindent The averages of the quadratic terms $\sim\hat{S}\hat{S}$
must be done separately as $\hat{S}$ does not commute with the probe
operators in general. It is convenient to introduce the matrix elements,\begin{eqnarray}
S_{ab}(m,n): & = & \langle\psi_{m}|\hat{S}|\psi_{n}\rangle\nonumber \\
 & = & \langle\psi_{m}|\mathcal{P}_{m}\hat{S}\mathcal{P}_{n}|\psi_{n}\rangle.\label{eq:-1}\end{eqnarray}
 In second-order perturbation theory, the generator reads \cite{Wolff},\begin{eqnarray}
S_{ab}(m,n) & = & \frac{J\alpha\langle\psi_{m}|\left(\boldsymbol{\tau}_{a}\cdot\boldsymbol{S}_{A}+\boldsymbol{\tau}_{b}\cdot\boldsymbol{S}_{B}\right)|\psi_{n}\rangle}{i(E_{m}-E_{n})}\nonumber \\
 & = & \frac{J\alpha}{i\Delta_{mn}}\left(\boldsymbol{\tau}_{a}\cdot\langle\psi_{m}|\boldsymbol{S}_{A}|\psi_{n}\rangle\right.\nonumber \\
 &  & \left.+\boldsymbol{\tau}_{b}\cdot\langle\psi_{m}|\boldsymbol{S}_{B}|\psi_{n}\rangle\right).\label{eq:-2}\end{eqnarray}

\noindent Inserting the resolution of the identity $\sum_{m}|\psi_{m}\rangle\langle\psi_{m}|$
in the expression for $C_{2}$, we get after some algebra,\begin{equation}
C_{2}=-\frac{4\alpha^{2}J^{2}}{3}\sum_{k>0}\sum_{\mu=x,y,z}\frac{|\langle\psi_{0}|\left(\boldsymbol{S}_{A}^{\mu}-\boldsymbol{S}_{B}^{\mu}\right)|\psi_{k}\rangle|^{2}}{\Delta_{k0}^{2}}\langle\boldsymbol{\tau}_{a}\cdot\boldsymbol{\tau}_{b}\rangle_{\textrm{can}}.\label{eq::App5-Canonical-17}\end{equation}
Indeed, we arrive at the desired result\begin{eqnarray}
\langle\boldsymbol{\tau}_{a}\cdot\boldsymbol{\tau}_{b}\rangle & = & \langle\boldsymbol{\tau}_{a}\cdot\boldsymbol{\tau}_{b}\rangle_{\textrm{can}}(1-\Phi)+\eta,\label{eq:App5-Canonical-18}\\
\Phi & = & \left(\frac{2\alpha J}{\sqrt{3}}\right)^{2}\sum_{k>0}\sum_{\mu=x,y,z}\frac{|\langle\psi_{0}|\left(\boldsymbol{S}_{A}^{\mu}-\boldsymbol{S}_{B}^{\mu}\right)|\psi_{k}\rangle|^{2}}{\Delta_{k0}^{2}}+\nonumber \\
 &  & +O(\alpha^{3}),\label{eq:App5-Canonical-19}\end{eqnarray}
 with $\langle\boldsymbol{\tau}_{a}\cdot\boldsymbol{\tau}_{b}\rangle_{\mathrm{can}}$
given previously in Eqs.~(\ref{eq:tau_tau_parameters})-(\ref{eq:tau_tau_can}),
and where $\eta=0$ up to second-order perturbation theory since no
constant term has emerged from our expansion. It can be shown that
this term is at most of fourth order, $O[(J\alpha/\Delta)^{4}]$ \cite{TeseAires}.

\section{Appendix 3 }

We have performed several careful checks of the results obtained from
the ALPS code. Our \textsf{looper} simulations used the following
Monte Carlo input parameters: number of thermalization steps, $15000$,
number of sweeps, $400000$. As a check, we used larger values of
these parameters with no change in the results. We have no evidence
whatsoever that our simulations failed to converge.

An important check of convergence is obtained through the measurement
of the staggered magnetization. In our simulation (for the square
lattice and $\alpha=0.05$), we obtained\begin{equation}
\sqrt{\langle\left(M_{s}^{z}\right)^{2}\rangle}=0.18008,\label{eq:}\end{equation}
 which gives a value $\sqrt{3}\sqrt{\langle\left(M_{s}^{z}\right)^{2}\rangle}\approx0.31191$
close to the known value of the sub-lattice magnetization in the thermodynamic
limit (our system is finite and has open boundary conditions which
may explain the small deviation).

The QMC error bars for the correlations are typically below $1\%$;
for instance, for the square lattice, $\alpha=0.05$, and the lowest
simulated temperature, we have an estimate from the simulations $\textrm{error}\left\{ \boldsymbol{\tau}_{a}\cdot\boldsymbol{\tau}_{b}/4\right\} \simeq3\left[\textrm{error\_MQ}\left\{ \langle\tau_{a}^{z}\tau_{b}^{z}\rangle/4\right\} \right]=0.000735,$
while the value of the correlation in this case is $\langle\boldsymbol{\tau}_{a}\cdot\boldsymbol{\tau}_{b}\rangle/4=-0.214852$.


\begin{thebibliography}{33}
\bibitem{DiVicenzo}D. P. DiVincenzo, Fortschritte der Physik \textbf{48},
771 (2000).

\bibitem{Cirac}J. I. Cirac and P. Zoller, Phys. Rev. Lett. \textbf{74},
4091 (1995).

\bibitem{DiVincenzo2}D. P. DiVincenzo \emph{et al.}, Nature (London)
\textbf{408}, 339 (2000).

\bibitem{SolidStateQC_withBus} X. Zhou \emph{et al.}, Phys. Rev.
Lett. \textbf{89}, 197903 (2002); S. C. Benjamin and S. Bose, Phys.
Rev. Lett. \textbf{90}, 247901 (2003); Man-Hong Yung, S. C. Benjamin
and S. Bose, Phys. Rev. Lett. \textbf{96}, 220501 (2006); Mark Friesen
\emph{et al.}, Phys. Rev. Lett. \textbf{98}, 230503 (2007).

\bibitem{Venuti}L. Campos Venuti, C. Degli Esposti Boschi, and M.
Roncaglia, Phys. Rev. Lett. \textbf{96}, 247206 (2006).

\bibitem{Osborne}T. J. Osborne and M. A. Nielsen, Phys. Rev. A \textbf{66},
032110 (2002).

\bibitem{Zell09}T. Zell, F. Queisser, and R. Klesse, Phys. Rev. Lett.
\textbf{102}, 160501 (2009).

\bibitem{DeChiara}G. D. Chiara, C. Brukner, R. Fazio, G. M. Palma
and V. Vedral, New. J. Phys. \textbf{8}, 95 (2006).

\bibitem{Review-Zureck-2003} W. H. Zureck, Rev. Mod. Phys. \textbf{75},
715 (2003).

\bibitem{2002Braun}D. Braun, Phys. Rev. Lett. \textbf{89}, 277901
(2002).

\bibitem{2003Benatti}F. Bennatti, R. Floreanini, and M. Piani, Phys.
Rev. Lett. \textbf{91}, 070402 (2003).

\bibitem{2008Paz}J. P. Paz and A. J. Roncaglia, Phys. Rev. Lett.
\textbf{100,} 220401 (2008).

\bibitem{2006Oh}S. Oh and J. Kim, Phys. Rev. A \textbf{73}, 062306
(2006).

\bibitem{1998Lidar}D. A. Lidar, I. L. Chung, and K. B. Whaley, Phys.
Rev. Lett. \textbf{81}, 2594 (1998).

\bibitem{2000Beige}A. Beige, D. Braun, B. Tregenna, and P. L. Knight,
Phys. Rev. Lett. \textbf{85}, 1762 (2000).

\bibitem{Venuti-b}L. Campos Venuti, C. Degli Esposti Boschi, and
M. Roncaglia, Phys. Rev. Lett. \textbf{99}, 060401 (2007).

\bibitem{Review-Manousakis}E. Manousakis, Rev. Mod. Phys. \textbf{63},
1 (1991);

\bibitem{CuGeO3}M. Hase, and I. Terasaki, and K. Uchinokura, Phys.
Rev. Lett. \textbf{70,} 3651 (1993).

\bibitem{1D-KCuF3}K. Hirakawa and Y. Kurogi, Prog. Theor. Phys. Suppl.
\textbf{46}, 147 (1970); H. Yoshizawa and K. Hirakawa, Phys. Rev.
B \textbf{21}, 2001 (1980).

\bibitem{AtomicChains}C. F. Hirjibehedin, \emph{et al.} Science \textbf{312},
1021 (2006).

\bibitem{Anderson} P. W. Anderson, \emph{Basic Notions of Condensed
Matter Physics} (Addison Wesley, New York, 1984).

\bibitem{negativity}G. Vidal, and R. F. Werner, Phys. Rev. A \textbf{65},
032314 (2002).

\bibitem{concurrence}W. K. Wootters, Phys. Rev. Lett. \textbf{80},
2245 (1998).

\bibitem{AFF_LDS}A. Ferreira and J. M. B. Lopes dos Santos, Phys.
Rev. A \textbf{77}, 034301 (2008).

\bibitem{VenutiPRA}L. Campos Venuti, S. M. Giampaolo, F. Illuminati,
P. Zanardi, Phys. Rev. A 76, 052328 (2007).

\bibitem{Book-Auerbach-2}A. Auerbach, \emph{Interacting Electrons
and Quantum Magnetism} (Springer, New York, 1994).

\bibitem{White}S. R. White, R. M. Noack, and D. J. Scalapino, Phys.
Rev. Lett. \textbf{73}, 886 (1994).

\bibitem{ALPS}H.G. Evertz, Adv. in Physics, \textbf{52}, 1 (2003);
S. Todo and K. Kato, Phys. Rev. Lett., \textbf{87}, 047203 (2001);
S. Todo in \emph{Condensed-Matter Physics}, edited by D.P. Landau,
S.P. Lewis, and H.-B. Schuettler (Springer, Berlin, 2003), Vol. 15;
F. Alet \emph{et al}., J. Phys. Soc. Jpn. Suppl. \textbf{74}, 30 (2005);
A. F. Albuquerque \emph{et al.,} J. Mag. Mag. Mat. \textbf{310}, 1187
(2007).

\bibitem{Chen}G. Chen, H-Q. Ding, and W. Goddard III, Phys. Rev.
B \textbf{46}, 2933 (1992).

\bibitem{Carlson}J. Carlson, Phys. Rev. B \textbf{40}, 846 (1989).

\bibitem{DiVincenzo3}D. Loss and D. P. DiVincenzo, Phys. Rev. A \textbf{57},
120 (1998); G. Burkard, D. Loss, and D. P. DiVincenzo, Phys. Rev.
B \textbf{59}, 2070 (1999).

\bibitem{Wolff} J. R. Schrieffer and P. A. Wolff, Phys. Rev. \textbf{149},
491 (1966).

\bibitem{TeseAires}Aires Ferreira\textbf{, }Ph.D. Thesis. \emph{The
Quantum-Classical Boundary: from Opto-Mechanics to Solid-State} (Universidade
do Porto, 2009\textbf{), }pre-print: arXiv: 0911.2217. 
\end{thebibliography}
\end{document}